\documentclass[journal]{IEEEtran}
\usepackage{amsmath}
\usepackage{booktabs}
\usepackage{array}
\usepackage{amssymb}
\usepackage{graphicx}
\usepackage{booktabs}
\usepackage{hyperref}
\ifCLASSINFOpdf
\else
\fi
\begin{document}
%

\title{Large Language Models for Solving Economic Dispatch Problem}

%
%
%

\author{Sina~Mohammadi,~\IEEEmembership{Graduate Student Member,~IEEE,}
        Ali~Hassan,~\IEEEmembership{Graduate Student Member,~IEEE,}
        Rouzbeh~Haghighi,~\IEEEmembership{Graduate Student Member,~IEEE,}
        Van-Hai~Bui,~\IEEEmembership{Senior Member,~IEEE,} and~Wencong~Su,~\IEEEmembership{Senior Member,~IEEE}
\thanks{S. Mohammadi, A. Hassan, R. Haghighi V.-H. Bui, and W. Su are with the Department of Electrical and Computer Engineering, University of Michigan– Dearborn, Dearborn, MI 48128, USA.}
}

\vspace{-20mm}

\maketitle

\vspace{-10mm}
\begin{abstract}

This paper investigates the capability of off-the-shelf large language models (LLMs) to solve the economic dispatch (ED) problem. ED is a hard-constrained optimization problem solved on a day-ahead timescale by grid operators to minimize electricity generation costs while accounting for physical and engineering constraints. Numerous approaches have been proposed, but these typically require either mathematical formulations, face convergence issues, or depend on extensive labeled data and training time. This work implements LLMs enhanced with reasoning capabilities to address the classic lossless ED problem. The proposed approach avoids the need for explicit mathematical formulations, does not suffer from convergence challenges, and requires neither labeled data nor extensive training. A few-shot learning technique is utilized in two different prompting contexts. The IEEE 118-bus system with 19 generation units serves as the evaluation benchmark. Results demonstrate that various prompting strategies enable LLMs to effectively solve the ED problem, offering a convenient and efficient alternative. Consequently, this approach presents a promising future solution for ED tasks, particularly when foundational power system models are available.

\end{abstract}


%
\IEEEpeerreviewmaketitle

\begin{table*}[ht]
\centering
\scriptsize
\caption{LLM Prompt}
\begin{tabular}{p{0.95\linewidth}}
\toprule
\textbf{Description of Problem} \\
\midrule
You are provided with a set of optimal generation dispatches (\(PG\)) for various loading scenarios, each associated with a specific total load demand (\(PD\)) and the corresponding minimum cost. The goal is to determine a new generation dispatch list for a total load demand, ensuring that the sum of the generation values equals the load demand while maintaining economic efficiency. \\
\midrule
\textbf{In-Context Examples (Population)} \\
\midrule
\(\displaystyle PD = 700 \text{ MW}, \quad \text{Cost} = 18077.53\) \\
\(\displaystyle PG = [50,\;10,\;20,\;40,\;5,\;5,\;30,\;10,\;50,\;20,\;40,\;80,\;100,\;60,\;2,\;50,\;108,\;10,\;10]\) \\
\\
\(\displaystyle PD = 2150 \text{ MW}, \quad \text{Cost} = 44448.51\) \\
\(\displaystyle PG = [59.42,\;10,\;20,\;485,\;5,\;20,\;223,\;10,\;85.55,\;195,\;40,\;80,\;100,\;71.85,\;2,\;70.18,\;653,\;10,\;10]\) \\
\\
\(\displaystyle PD = 3600 \text{ MW}, \quad \text{Cost} = 81779.65\) \\
\(\displaystyle PG = [505,\;10,\;20,\;485,\;17,\;20,\;223,\;16.85,\;308,\;195,\;40,\;80,\;100,\;453.41,\;2,\;451.74,\;653,\;10,\;10]\) \\
\\
\(\displaystyle PD = 5050 \text{ MW}, \quad \text{Cost} = 127038.67\) \\
\(\displaystyle PG = [505,\;10,\;221,\;485,\;17,\;20,\;223,\;53,\;308,\;195,\;45.41,\;530.98,\;503.42,\;509,\;10,\;637,\;653,\;108,\;16.19]\) \\
\\
\(\displaystyle PD = 6500 \text{ MW}, \quad \text{Cost} = 189132.65\) \\
\(\displaystyle PG = [505,\;70,\;221,\;485,\;17,\;20,\;223,\;53,\;308,\;195,\;441,\;784,\;1182,\;509,\;10,\;637,\;653,\;108,\;79]\) \\
\midrule
\textbf{Non-Evolutionary Algorithm Task Instruction} \\
\midrule
Generate a new list of generation dispatches \(PG\) for a total load demand \(PD = 727\) MW. The solution should follow the trend observed in the given data, maintaining proportionality and logical scaling of generator contributions with minimum cost value. \\
\midrule
\textbf{Evolutionary Algorithm Task Instruction} \\
\midrule
\begin{enumerate}
    \item Choose two dispatch scenarios from the provided data. These sets serve as parent solutions for generating a new candidate.
    \item Combine elements from the two selected parent dispatches to form a new candidate dispatch.
    \item Mutate the candidate dispatch obtained from the crossover.
    \item Repeat the selection, crossover, and mutation steps until you generate 10 candidate dispatch sets.
    \item Evaluate these 10 candidates based on their estimated cost, then select the best solution and provide its vector form.
\end{enumerate}
\textbf{Note:} Do not include any code; ensure the solution maintains exact power balance and respects the observed generator limits.
\\
\bottomrule
\end{tabular}
\label{prompt structure}
\vspace{-5mm}
\end{table*}

\section{Introduction}

\IEEEPARstart{I}{n} power systems, economic dispatch (ED) is an optimization problem that involves allocating electricity generation to meet system demand while minimizing total operating costs. The assignment of generation amounts to each unit is determined based on the cost coefficients of the individual generators, aiming to achieve the most economically efficient distribution of power production across the system. This optimization problem is fundamental for power system operations and has significant implications for the economic viability of power utilities and the overall efficiency of electricity markets \cite{Chowdhury1990}. The optimal generation must meet the load demand along with several constraints, such as line flow limits, generation capacity limits, ramp rate limits, and power balance \cite{nappu2014transmission}. ED is performed one day earlier for the day-ahead electricity market \cite{sun2020day} and also more frequently (e.g., 5 minutes) for the real-time market \cite{reddy2015real}. Traditionally, ED is performed using numerical methods such as the gradient method or Newton's method \cite{qin2019newton}, which are the iterative methods based on calculating the gradient of the objective function until the solution converges. The number of steps and time it takes to reach a solution depends on the length of the problem and initial guess \cite{saadat2010power}.  Optimization methods such as linear programming (LP) \cite{nabona1973optimisation} and mixed integer linear programming (MILP) \cite{nemati2018optimization} linearize the cost function of the generating units and solve the resulting solution to minimize the generation cost. Quadratic programming \cite{reid1973economic}, on the other hand, allows the quadratic terms in the cost function while keeping the constraints linear. Heuristic and metaheuristic techniques, such as genetic algorithm (GA) \cite{ponciroli2020improved}, particle swarm optimization (PSO)  \cite{basu2015modified}, and simulated annealing (SA) \cite{aghaee2016economic}, rely on gradual improvement in the solution using nature-based algorithms. Artificial intelligence (AI) techniques such as artificial neural networks (ANN) \cite{kim2020neural}, fuzzy logic \cite{nasiruzzaman2008implementation}, and reinforcement learning (RL) \cite{lin2020deep}, are also employed to solve the ED problem. These methods either rely on the large training dataset or huge environment interaction. This data is then used to train the AI model, which can subsequently predict solutions for unseen load scenarios based on its training. The AI approach offers the potential to capture complex system behaviors and adapt to changing conditions in the power system, making it an increasingly valuable tool to address modern challenges in ED.

In recent years, LLMs have ushered in a new era of AI technologies, reinforced by natural language capabilities and functioning as foundational models with which anyone can interact and chat. Initially, the abilities of LLMs were explored only in text analysis, such as writing a poem, composing an email, or engaging in simple chat interactions about general topics. However, updated versions of LLMs have made the integration of natural language processing (NLP) in the technical domain possible. Today, multiple LLMs have been developed with varying task-handling capabilities, and their potential to address different engineering tasks is currently being explored and studied. Among these applications, using LLMs to solve optimization problems has recently gained significant interest. In \cite{huang2024large}, the authors explored LLMs for security-constrained unit commitment and the EV charging scheduling problem. However, to our knowledge, LLMs have not been explored for the ED problems of power systems. This paper presents an ED problem framed for the IEEE 118 bus system, which is used as a prompt for the  latest reasoning models of standard LLMs.  Besides, this paper is the first work which investigates solving a hard constrained optimization problem for power systems in only one interaction with LLM.

The main contributions of this paper are as follows :




\begin{enumerate}
    \item This work explores two state-of-the-art large language models (LLMs), namely O3-mini-high (O3 mini and O1, i.e., ChatGPT) and R1 (DeepSeek), for solving the economic dispatch (ED) problem in power systems using both non-evolutionary and evolutionary algorithms with few-shot prompting.
    \item Unlike previous approaches that relied on meta prompting and multiple interactions, the proposed method requires only a single interaction to solve the ED problem.
    \item Inequality constraints for generation units are incorporated into the few-shot samples, allowing the LLM to learn how to find an ED solution with minimal constraint violations even without explicit generation limits. Moreover, cost coefficients are omitted from the prompt to reduce its length and enhance the model's exploration capabilities.
\end{enumerate}

\section{Methodology}
\subsection{LLMs for Solving Optimization Problems}

Google DeepMind published a paper that extensively explored the capabilities of LLMs in solving various benchmark optimization problems \cite{yang2024largelanguagemodelsoptimizers}. Optimization by prompting (OPRO) is the general concept of the paper, which employs a feedback loop that scores the LLM responses to evaluate and generate more accurate problem solutions through a sequence of prompts and responses. This method generally uses a few-shot learning technique by providing some sample examples as initial information or ground truth for the LLM model. Following this work, evolutionary optimization using LLMs was introduced \cite{liu2024large}. In this approach, the LLM model mimics the evolutionary genetic algorithm concept to solve optimization problems. As stated in \cite{liu2024large}, the proposed method outperforms the OPRO method developed by Google DeepMind for benchmark optimization problems such as the traveling salesman problem.
Inspired by these works, the ability of multiple off-the-shelf LLMs to solve the ED problem is studied in this paper. Few-shot learning without a scoring function and evolutionary prompt optimization are utilized to address the ED problem, which is a crucial optimization challenge in power system studies. These two methods are referred to as non-evolutionary and evolutionary prompting approaches in this paper.

\vspace{-5mm}
\subsection{LLMs for Solving ED Problem}

The ED is a fundamental optimization problem that is solved to minimize the cost of generation in day-ahead power system planning. The classic ED includes a quadratic cost function that represents each generation unit's contribution to the total cost, power balance as the equality constraint, and generation limits as the inequality constraints. The general formulation of the classic lossless ED is as follows:
\begin{align*}{}
    \text{min}   & \sum_{i=1}^{N} \left( a_i P_{G_i}^2 + b_i P_{G_i} + c_i \right) \\
    \text{subject to:} & \quad \sum_{i=1}^{N} P_{G_i} = P_D \\
                       & \quad P_{G_i}^{\min} \leq P_{G_i} \leq P_{G_i}^{\max},\quad \forall i \in \{1,\dots,N\}
\end{align*}
where a, b and c are cost coefficients, N is the number of generation units, $P_{G}$ is the generation unit value and $P_{D}$ is the total demand in the power system.
In this paper, the LLM is used to solve the ED problem by utilizing two different prompting approaches. In the first scenario, a few-shot learning approach, referred to as the non-evolutionary approach, is introduced, where minimum information is provided as the prompt for the LLM model. In this approach, the few-shot samples are carefully selected to cover a wide range of possible scenarios for demand variations. The minimum and maximum loading limits can be derived based on the base load and the maximum generation limits of the generation units. Additionally, the cost coefficients and generation limits are not directly included in the prompt. This reduces the prompt length and allows the LLM to search for the ED solution with more exploration capability without considering numerical content.
In the second approach, the evolutionary concept is employed to solve the ED problem. Similar to the first scenario, a few samples for different loading scenarios, along with their associated cost values, are provided as pre-solved ground truth data using traditional solvers. Then, the evolutionary algorithm is applied as the LLM's task for solving ED. For this task, two sets of provided samples are first selected as candidates. Next, crossover and mutation are performed for a limited number of iterations based on the newly generated dispatch sets until the optimal candidate is found. In this approach, an evolutionary guideline is defined for the LLM to search for the ED solution, which can be categorized as a semi-supervised method. This approach is inherently different from the first approach, where the LLM relies solely on its own search mechanism and reasoning to find the solution to the problem. 

\begin{table}[t]
\centering
\scriptsize
\caption{IEEE 118-Bus Data Required for ED}
\label{118 data}
\begin{tabular}{cccccc}
\toprule
Bus & \(P_{\min}\) & \(P_{\max}\) & \(a\) & \(b\) & \(c\) \\
\midrule
10  & 50  & 505   & 0.00043 & 24.98  & 500 \\
12  & 10  & 85    & 0.00194 & 124.58 & 300 \\
25  & 20  & 221   & 0.00254 & 28.95  & 100 \\
26  & 40  & 485   & 0.001 & 22.22  & 50 \\
31  & 5  & 17    & 0.002 & 25.99  & 350 \\
46  & 5  & 20    & 0.0005 & 24.20  & 100 \\
49  & 30  & 223   & 0.0009 & 16.67  & 100 \\
54  & 10  & 53    & 0.0012 & 27.28  & 100 \\
59  & 50  & 308   & 0.001 & 24.86  & 20 \\
61  & 20  & 195   & 0.005 & 16.06  & 50 \\
65  & 40  & 441   & 0.006 & 34.78  & 100 \\
66  & 80  & 784   & 0.0025 & 32.67  & 90 \\
69  & 100  & 1182  & 0.0095 & 25.76  & 300 \\
80  & 60  & 509   & 0.003 & 24.60  & 200 \\
87  & 2  & 10    & 0.004 & 34.07  & 100 \\
89  & 50  & 637   & 0.003 & 24.61  & 100 \\
100 & 60  & 653   & 0.002 & 12.61  & 40 \\
103 & 10  & 108   & 0.0016 & 28.65  & 50 \\
111 & 10  & 79    & 0.0088 & 35.04  & 80 \\
\bottomrule
\end{tabular}
\vspace{-5mm}
\end{table}

Table \ref{prompt structure} illustrates the prompt structure (the loading scenario can be changed for other sceanarios) for one of the scenarios considerd in the paper for both the first and second approaches considered in this paper. As shown, the ED problem is described in simple terms using natural language without including any mathematical formulations. This simplifies the expression of the ED problem and makes it more accessible for grid operators, especially if foundational models are developed for grid monitoring and control purposes in the near future.

\section{Results}

This section presents the performance analysis of different LLMs in solving the ED problem, considering the two approaches discussed in the previous section. In this paper, the IEEE 118-bus test system is considered as the case study, with 19 generation units. The required data for the ED of these test systems are obtained from PGLib-OPF, an open-source database for benchmark power grids, which includes the generation limits, cost coefficients, and initial demand values of the systems \cite{babaeinejadsarookolaee2021powergridlibrarybenchmarking}. However, this data introduces a linear cost term with only b values. To have a quadratic cost function the a and c values are added as seen in Table \ref{118 data}.


\begin{figure}[t]
\centering
\begin{tabular}{c}
\includegraphics[width=0.4\textwidth]{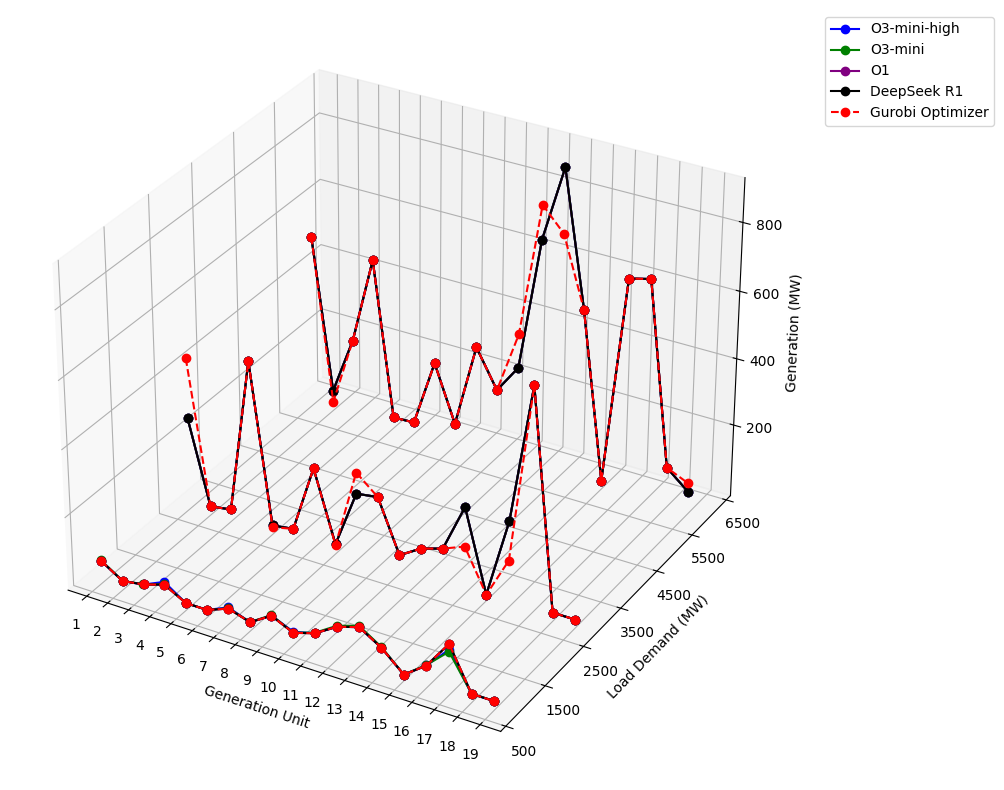}\\
(a)\\
\includegraphics[width=0.4\textwidth]{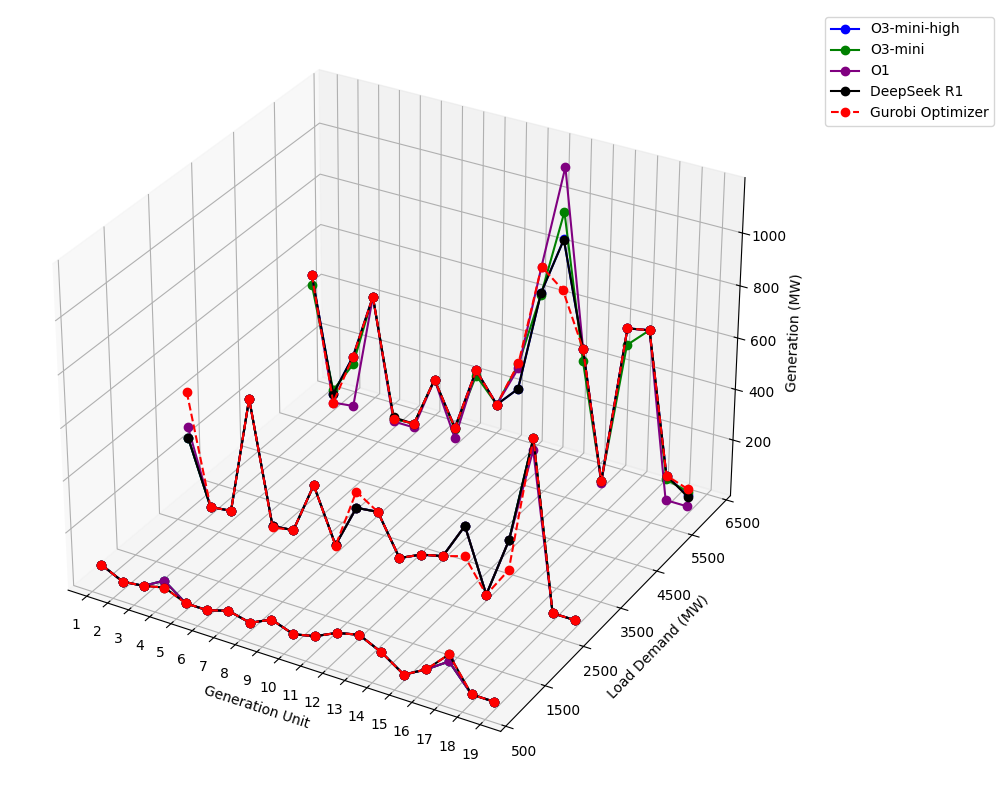}\\
(b)\\
\end{tabular}
\caption{Non-Evolutionary and evolutionary ED results for selected loading scenarios: LLMs versus Gurobi (a) Non-Evolutionary prompt (b) Evolutionary prompt}
\label{pg trends}
\vspace{-5mm}
\end{figure}

In both prompting approaches used to solve the ED problem, Gurobi Optimizer is deployed to solve ED for five different samples. These five samples are selected to represent a well-distributed range of possible loading scenarios bounded by the base load and maximum generation capacity of the units. These five samples serve as the few-shot examples provided to the LLM model to help it learn how decision variables change in the ED problem. In other words, the few-shot samples encapsulate all necessary information for finding the ED solution. Besides, this approach can further challenge the LLMs for predicting the ED solutions which is more desired in this work. In this study, the newest available LLM versions—GPT O3-mini-high, O1, and O3-mini—along with DeepSeek R1 are considered for prompting. These models are reinforced with reasoning capabilities and have demonstrated better performance on various tasks compared to older versions, particularly in math-related questions. In addition, these methods can better approach finding the nonlinear relationships among few-shot samples for finding the ED problem solution.

\begin{figure}[t]
\centering
\begin{tabular}{c}
\includegraphics[width=0.48\textwidth]{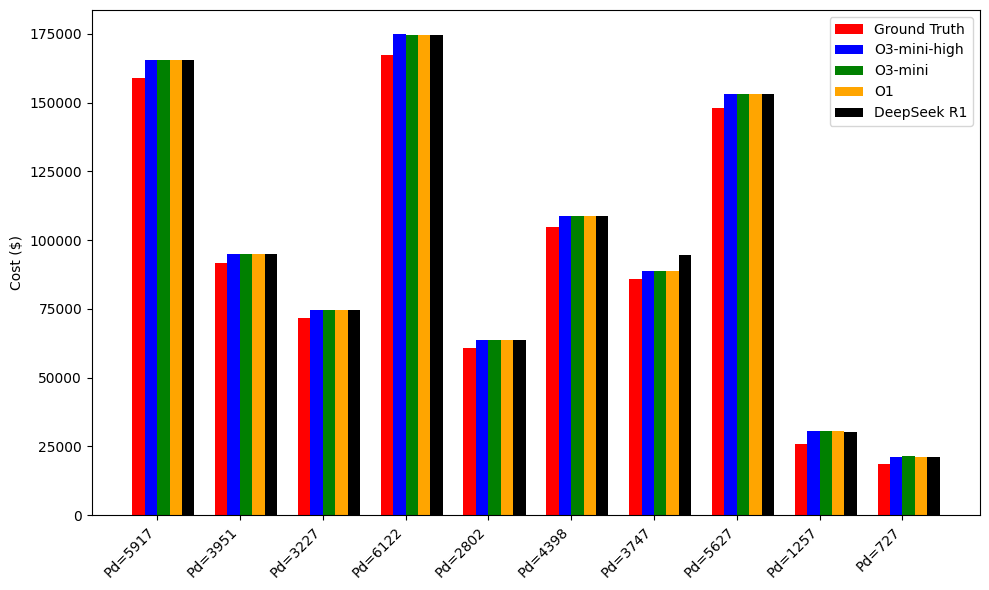}\\
(a)\\
\includegraphics[width=0.48\textwidth]{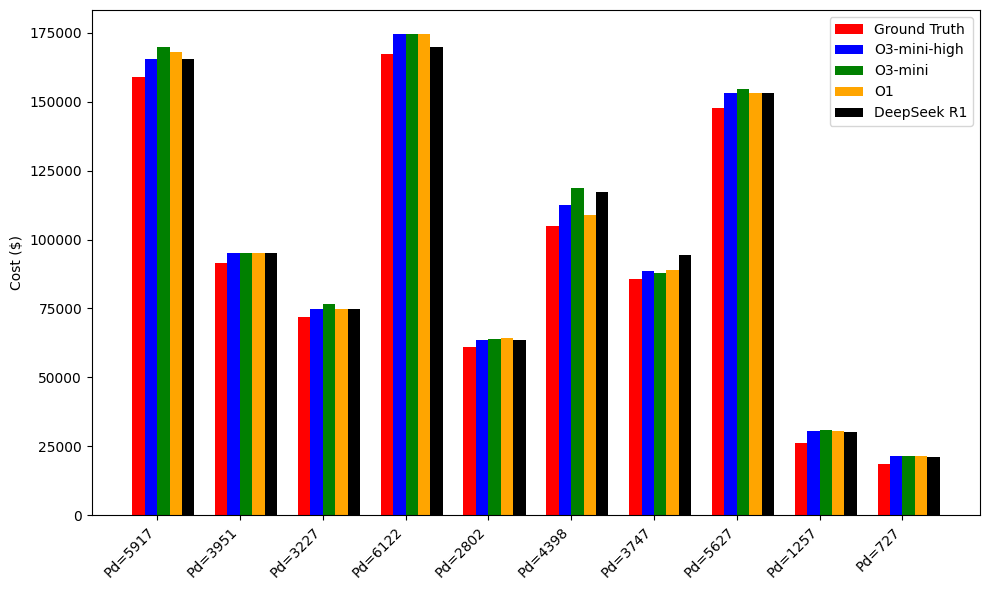}\\
(b)\\
\end{tabular}
\caption{Cost comparison results for ED solutions considering both approaches. (a) non-evolutionary algorithm (b)evolutionary algorithm}
\label{cost comparison}
\vspace{-3mm}
\end{figure}


To test the proposed approaches, 10 new loading scenarios, different from the 5 few-shot samples, are considered to evaluate the performance of different LLMs based on the prompt structures shown in Table I. The number of loading scenarios could be selected as a few hundred, but the API key for reasoning models is too expensive, and the authors use the web-based LLM edition for generating ED solutions. Additionally, each response generation takes a few minutes, depending on which LLM is selected, and the DeepSeek R1 model experiences server busy issues after only a few chats. Thus, only 10 loading scenarios are selected for solving ED via LLMs in this paper. The loading scenarios are selected randomly within the minimum and maximum demand range and are $P_d = [727, 1257, 2802, 3227, 3747, 3951, 4398, 5627, 5917, 6122]$ MW. This random pattern ensures the maximum level of difficulty for LLMs in solving ED. Fig. \ref{pg trends} displays the predictions of different LLM models for the non-evolutionary and evolutionary algorithms. As evident, all LLM models can provide a generation pattern similar to the ground truth obtained from the Gurobi Optimizer. It is worth mentioning that the prediction results do not need to match the ground truth numbers exactly, as some generation units may have different values compared to the Gurobi results, provided that the cost is close and there is minimum violation of power balance.



Fig. \ref{cost comparison} show the cost comparisons for both non-evolutionary and evolutionary algorithms. Based on the results, it is evident that in the non-evolutionary algorithm, as expected, the randomness is higher, and the error can vary. Additionally, different models exhibit different errors depending on the loading scenario (see Table \ref{tab:relative_errors1} and \ref{tab:relative_errors2}), making it impossible to identify the best and worst methods. However, as shown in Table \ref{tab:relative_errors1} and \ref{tab:relative_errors2}, all models demonstrate lower accuracy for $P_d = 1257$ MW and $P_d = 727$ MW. One reason for this is the boundary sensitivity of LLMs. Based on the experiments, LLMs use only the linear scaling method for predicting the ED solutions for load scenarios that are very close to the few-shot provided ground truth, which leads to higher errors without considering the rate of change in the other provided samples. 

\begin{figure}[t]
\centering
\begin{tabular}{c}
\includegraphics[width=0.4\textwidth]{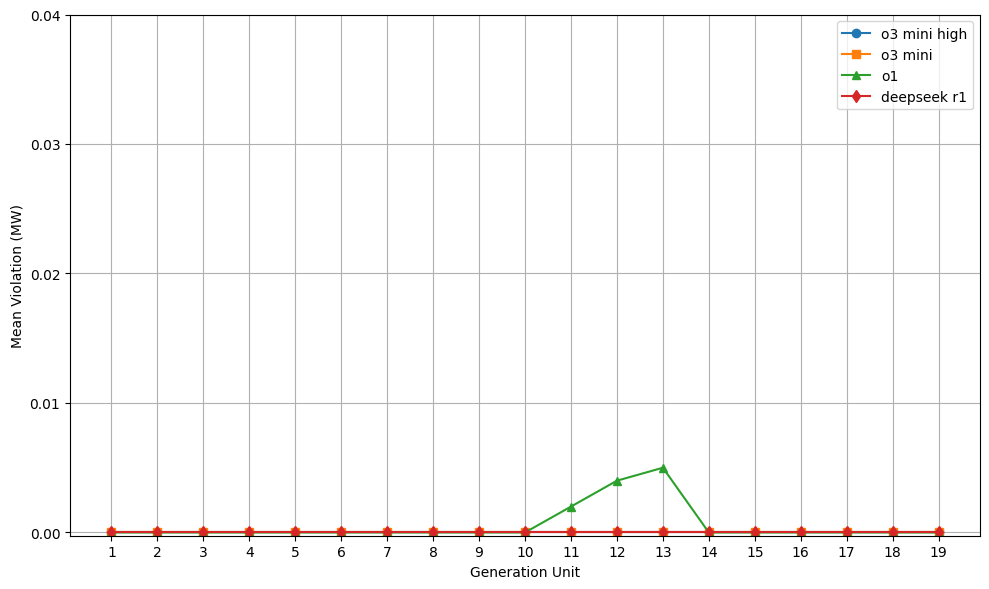}\\
(a)\\
\includegraphics[width=0.4\textwidth]{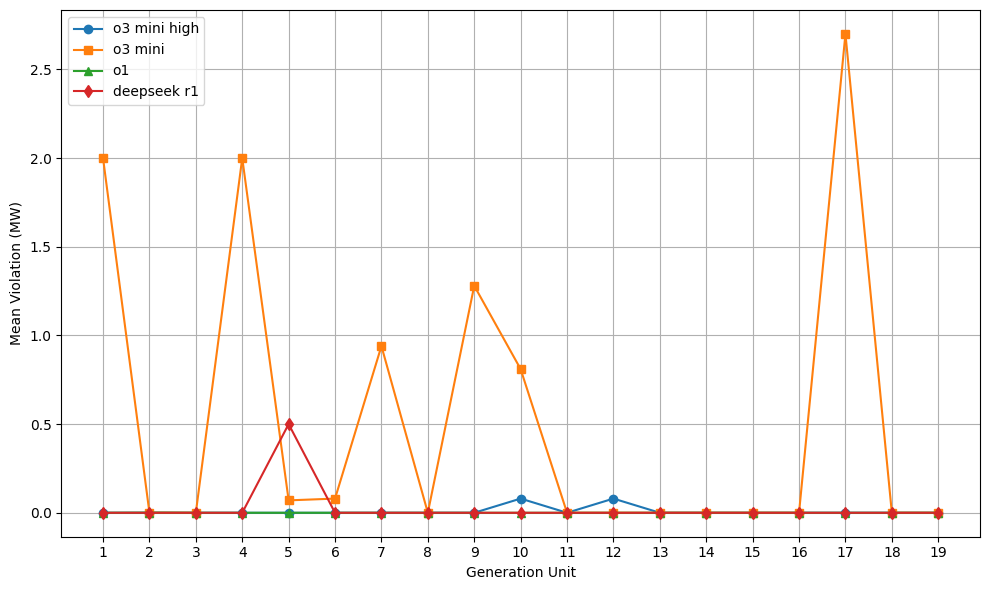}\\
(b)\\
\includegraphics[width=0.4\textwidth]{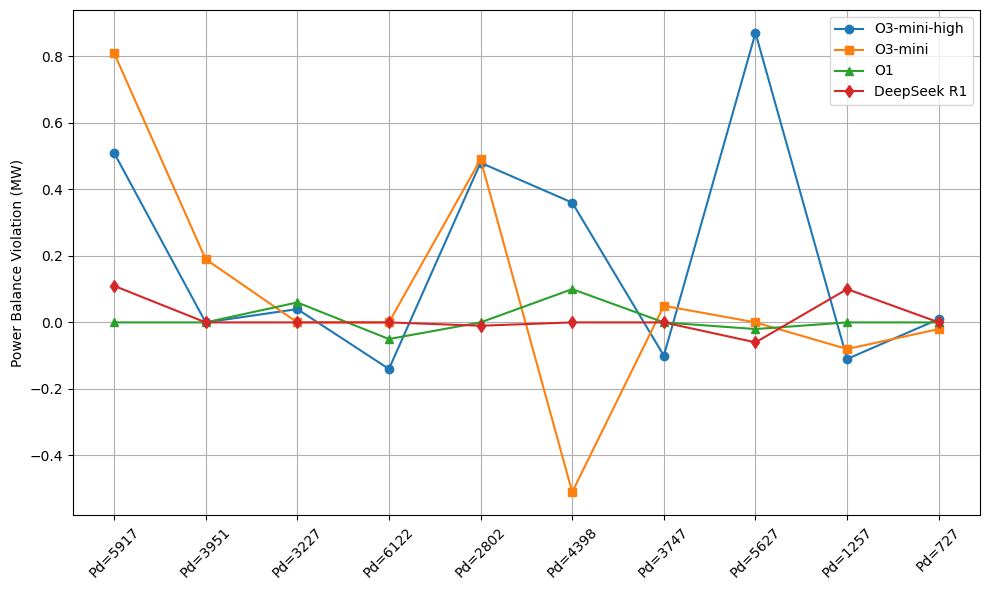}\\
(c)\\
\includegraphics[width=0.4\textwidth]{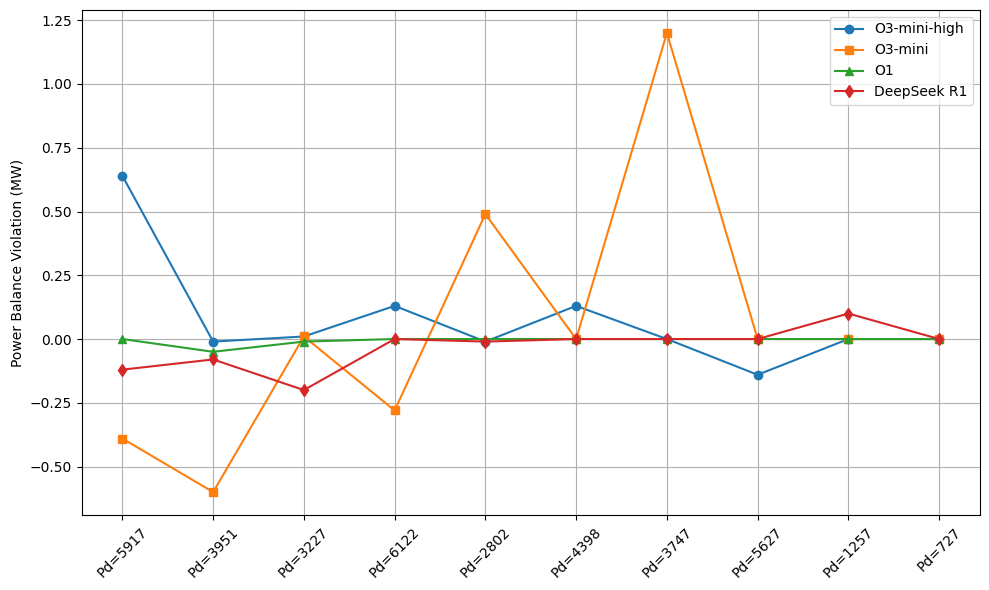}\\
(d)\\
\end{tabular}
\caption{Generation limitation and power balance violation for different loading scenarios considering different LLMs. (a) non-evolutionary algorithm generation violation (b) evolutionary algorithm generation violation (c) non-evolutionary algorithm power balance violation (d) evolutionary algorithm power balance violation}
\label{violations}
\vspace{-5mm}
\end{figure}

Another important factor is that the desired solution for ED should minimize constraint violations. Fig. \ref{violations} shows the mean values of violations for generation and power balance across different LLMs. As shown, in terms of generation violations, the non-evolutionary algorithm performs better with almost zero violations. For power balance, the violations are higher, whereas the O1 and DeepSeek R1 models demonstrate better performance with nearly zero violations for different random loading scenarios.

\begin{table}[t]
\centering
\scriptsize
\caption{Generation Cost Relative Error for the Evolutionary Approach (in \%)} 
\begin{tabular}{lcccc}
\hline
Load Demand & O3-mini-high & O3-mini & O1 & DeepSeek R1 \\
\hline
5917 & 4.16 & 4.17 & \textbf{4.15} & \textbf{4.15} \\
3951 & \textbf{3.81} & 3.82 & 3.81 & \textbf{3.81} \\
3227 & \textbf{4.07} & \textbf{4.07} & 4.08 & 4.07 \\
6122 & 4.56 & \textbf{4.38} & 4.39 & 4.39 \\
2802 & 4.68 & 4.67 & 4.67 & \textbf{4.66} \\
4398 & 3.84 & \textbf{3.82} & 3.83 & \textbf{3.82} \\
3747 & \textbf{3.58} & \textbf{3.58} & \textbf{3.58} & 10.17 \\
5627 & 3.60 & 3.58 & 3.58 & \textbf{3.57} \\
1257 & 18.20 & 18.20 & 18.23 & \textbf{16.56} \\
727  & 15.46 & 16.71 & \textbf{14.80} & \textbf{14.80} \\
\hline
\end{tabular}
\label{tab:relative_errors1}
\end{table}

\begin{table}[t]
\centering
\scriptsize
\caption{Generation Cost Relative Error for the Non-Evolutionary Approach (in \%)}
\begin{tabular}{lcccc}
\hline
Load Demand & O3-mini-high & O3-mini & O1 & DeepSeek R1 \\
\hline
5917  & 4.16   & 6.86   & 5.65   & \textbf{4.11} \\
3951  & 3.81   & \textbf{3.80}  & 3.81   & 3.81 \\
3227  & \textbf{4.07}   & 6.75   & \textbf{4.07}  & 4.20 \\
6122  & 4.39   & 4.35   & 3.79   & \textbf{1.63} \\
2802  & \textbf{4.66}  & 4.67   & 5.41   & \textbf{4.66} \\
4398  & 7.33   & 13.30  & \textbf{3.75}  & 11.85 \\
3747  & 3.23   & 2.54   & \textbf{0.35}  & 10.17 \\
5627  & \textbf{3.58}   & 4.65   & 3.70   & \textbf{3.58} \\
1257  & 18.11  & 18.21  & 17.21  & \textbf{16.56} \\
727   & 16.17  & 16.17  & 16.17  & \textbf{14.81} \\
\hline
\end{tabular}
\label{tab:relative_errors2}
\vspace{-5mm}
\end{table}

\vspace{-4mm}
\section{Conclusions}

In this paper, the capabilities of LLMs for solving the ED problem are explored as the first attempt to address a hard-constrained optimization problem in the power system domain with only single shot interaction with LLM. Based on the results, LLMs can find the ED solution with a good accuracy by employing a few-shot learning process in the prompt without any repetitive procedures. Two evolutionary and non-evolutionary algorithms were tested, which can be considered as non-guided and semi-guided schemes, respectively, to evaluate the performance of multiple LLM models in solving the ED problem. The non-evolutionary approach relies solely on the LLM's self-reasoning to find the solution, while the evolutionary algorithm performs an iterative search by generating new sets of possible answers for the problem. The results show that there is no big difference between the two approaches taken for solving ED. However, the non-evolutionary algorihm is more accurate in terms of lower constraint violations. This indicates that the evolutionary model does not inherently provide better exploration capabilities for an LLM model when addressing ED problem. In the future, the ability of LLMs to solve more complex optimization problems, such as DC/AC optimal power flow, will be investigated.

\vspace{-2mm}
\bibliography{Bibliography}
\bibliographystyle{ieeetr}

%

\end{document}